\documentclass[12pt]{article}
\pdfoutput=1

\usepackage{graphicx}
\usepackage{wrapfig}
\usepackage{amsfonts}

\input epsf.sty
\def\timesbox{\hbox{$\scriptscriptstyle\times$}}
\def\ant{ {{\lower 1ex  \timesbox} \atop {\raise 1.5ex  \timesbox}}}
\newcommand\ZZZ{{\hbox{ Z\kern-1.6mm Z}}}
\newcommand{\Iop}{\relax{\rm I\kern-.18em I}}
\newcommand{\Lop}{\relax{\rm I\kern-.18em L}}
\newcommand{\dop}{\relax{\rm I\kern-.8em d}}
\newcommand{\one}{{\hbox{ 1\kern-1.2mm l}}}

\newcommand{\beq}{\begin{equation}}
\newcommand{\eeq}{\end{equation}}
\newcommand{\bea}{\begin{eqnarray}}
\newcommand{\eea}{\end{eqnarray}}
\newcommand{\ra}{\rangle}
\newcommand{\la}{\langle}
\newcommand{\lt}{\left}
\newcommand{\rt}{\right}

\newcommand{\del}{\partial}

\newcommand{\dlt}{\delta}
\newcommand{\eps}{\epsilon}

\newcommand{\s}{\sigma}

\newcommand{\Dlt}{\Delta}

\newcommand{\cD}{{\cal D}}

\newcommand{\cF}{{\cal F}}
\newcommand{\cH}{{\cal H}}

\newcommand{\cL}{{\cal L}}
\newcommand{\cLM}{{\cal LM}}
\newcommand{\cM}{{\cal M}}

\newcommand{\cT}{{\cal T}}
\newcommand{\cU}{{\cal U}}

\newcommand{\ha}{\hbox{a}}
\newcommand{\hb}{\hbox{b}}

\newcommand{\hd}{\hbox{d}}
\newcommand{\he}{\hbox{e}}

\newcommand{\mD}{\mathbb{D}}

\newcommand{\mR}{\mathbb{R}}

\newcommand{\mZ}{\mathbb{Z}}

\newcommand{\hbd}{{\hbox{\bf d}}}

\begin{document}

{}~
{}~

\vskip 2cm

\centerline{\Large \bf On a semi-classical limit of loop space quantum}
\centerline{\Large \bf mechanics}

\medskip

\vspace*{4.0ex}

\centerline{\large \rm Partha Mukhopadhyay }

\vspace*{4.0ex}

\centerline{\large \it The Institute of Mathematical Sciences}
\centerline{\large \it C.I.T. Campus, Taramani}
\centerline{\large \it Chennai 600113, India}

\medskip

\centerline{E-mail: parthamu@imsc.res.in}

\vspace*{5.0ex}

\centerline{\bf Abstract}
\bigskip

Following earlier work, we view two dimensional non-linear sigma model with target space $\cM$ as a single particle relativistic quantum mechanics in the corresponding free loop space $\cLM$. In a natural semi-classical limit ($\hbar=\alpha' \to 0$) of this model the wavefunction localizes on the submanifold of vanishing loops which is isomorphic to $\cM$. One would expect that the relevant semi-classical expansion should be related to the tubular expansion of the theory around the submanifold and an effective dynamics on the submanifold is obtainable using Born-Oppenheimer approximation. In this work we develop a framework to carry out such an analysis at the leading order in $\alpha'$-expansion. In particular, we show that the linearized tachyon effective equation is correctly reproduced up to divergent terms all proportional to the Ricci scalar of $\cM$. The steps leading to this result are as follows: first we define a finite dimensional analogue of the loop space quantum mechanics (LSQM) where we discuss its tubular expansion and how that is related to a semi-classical expansion of the Hamiltonian. Then we study an explicit construction of the relevant tubular neighborhood in $\cLM$ using exponential maps. Such a tubular geometry is obtained from a Riemannian structure on the tangent bundle of $\cM$ which views the zero-section as a submanifold admitting a tubular neighborhood. Using this result and exploiting an analogy with the toy model we arrive at the final result for LSQM.

\newpage

\tableofcontents

\baselineskip=18pt

\section{Introduction and summary}
\label{s:intro}

Strings in curved background is a well studied problem \cite{review}. Usually the semi-classical expansion is formulated using the background field method of quantum field theory (QFT) in Lagrangian framework \cite{honerkamp71, friedan80, gaume81}. An attractive feature of this formulation is the use of Riemann normal coordinate (RNC) \cite{eisenhart} expansion. This enables one to keep the Riemannian structure of the target manifold $\cM$ manifest. 

Although it is not usually used for QFT computations, Hamiltonian framework, on the other hand, is conceptually appealing. It is natural to view the two dimensional non-linear sigma model (NLSM) under consideration as a single particle relativistic quantum mechanics in the infinite dimensional free loop space $\cLM$ corresponding to $\cM$ \cite{witten82, witten87, schreiber, frenkel}. In \cite{dwv} we discussed a framework of describing this quantum mechanics, hereafter called loop space quantum mechanics (LSQM), for the bosonic sigma model in terms of general coordinates in $\cLM$ keeping the infinite dimensional Riemannian structure manifest. This may be viewed as a formal $\hbar$-deformation of the classical theory as divergences are present in the form of infinite dimensional traces. The problem of regularizing these divergences was emphasized earlier in \cite{witten82, freund, witten87}. As a first step towards this direction, in this article we discuss a semi-classical limit of LSQM and motivate the use of Fermi normal coordinate (FNC) \cite{florides, tubes} expansion describing the tubular neighborhood of $\cM$ when it is viewed as the submanifold of vanishing loops embedded in $\cLM$.\footnote{The view of studying $\cM$ through its embedding in $\cLM$ was considered earlier by E. Witten in \cite{witten87}.}  

We now roughly describe the general idea. One expects that in $\hbar = \alpha'  \to 0$ limit the worldsheet theory should reduce to a theory of particles in $\cM$. One also notices that LSQM has a potential which minimizes to zero on the submanifold of zero loops. Therefore a natural semi-classical limit is given by the situation where the wavefunction localizes on this submanifold. The general idea is to use Born-Oppenheimer type approximation to adiabatically decouple the longitudinal ({\it slow}) and transverse ({\it fast}) degrees of freedom\footnote{This is similar in spirit to the discussion of degenerate Morse theory in \cite{witten82}. As explained below, we will study this problem in more detail following certain other literature.} and finally to compute the effective theory on $\cM \hookrightarrow \cLM$ order by order in $\hbar$. 

A complete understanding of the above procedure requires several technical questions to be answered. Some of them are as follows.
\begin{enumerate}
\item
How to perform tubular expansion of tensors around a submanifold embedded in a higher dimensional ambient space?
\item
Given a suitable quantum mechanical problem in the ambient space, how to set up the relevant semi-classical expansion of the Hamiltonian which relates to the above tubular expansion? How to get an effective theory on the submanifold? 
\item
Given the understanding of the above questions in a finite dimensional case, how to apply them to our present context of loop space? 
\end{enumerate} 

We will discuss all the above three topics successively and our final goal will be to derive the linearized effective equation for the tachyon fluctuation at leading order in $\alpha'$-expansion. We will show that our analysis correctly reproduces the known result up to divergent terms all proportional to the Ricci scalar of $\cM$. Below we briefly discuss these topics to indicate how this result will be arrived at. This will also clarify relation to other works in the literature. 

The first question is discussed in \S \ref{s:FNC} (and in appendix \ref{a:vielbein}). Here we explain our basic set up for a finite dimensional submanifold embedding, introduce FNC and review the results of \cite{tubular}. In \cite{tubular}, by generalizing the techniques of \cite{muller}, we find all order FNC-expansion of vielbein components in the neighborhhood of a submanifold (say $M$) embedded in a pseudo-Riemannian ambient space (say $L$)\footnote{$M$ and $L$ are our finite dimensional analogues of $\cM$ and $\cLM$ respectively.}. The expansion coefficients are given by certain tensors of $L$, all evaluated at $M \hookrightarrow L$. For vielbein these tensors are given by
combinations of various powers of the curvature, their covariant derivatives and spin connection. For the rest of our analysis the FNC-expansion of the metric tensor up to quadratic order, as given in eq.(\ref{g-expansion}), will be crucially used.

To address the second question we consider a finite dimensional analogue of LSQM in \S \ref{s:finite-dim}.  The analysis in this section is along the line of what is usually known as {\it constrained quantum system} in the literature. A partial list of references is \cite{marcus, maraner, mitchell, tenuta, wachsmuth}. Here one considers a non-relativistic classical system in an ambient space with a potential that tries to confine the motion into a submanifold. The idea is to realize this constraint at the quantum mechanical level through localization of wavefunction. This is done by rescaling the model with certain tunable parameter (e.g. representing the strength of the restoring force) in such a way that makes the transverse directions {\it fast} in the Born-Oppenheimer sense when the parameter is small. In our case the tunable parameter is the scale $\hbar$ and therefore the procedure gives a semi-classical expansion of the theory. In \S \ref{s:finite-dim} we give precise definition of the potential of our model and the procedure leading to semi-classical expansion of the Hamiltonian. This shows how the contribution at a given order in $\hbar$ is related to tubular expansion of various geometric quantities at different orders. Finally, we define and compute an analogue of linearized tachyon effective equation at leading order in $\hbar$-expansion within this toy model.

The usefulness of this study lies in the fact that it is free of divergences. Moreover, as hinted in the next paragraph, there exists an analogy which can be exploited to translate the end results of the toy model to the case of LSQM. Once this is done it exhibits the pattern of divergences that are expected in the actual LSQM computations. This is how we arrive at the final result for the tachyon effective equation as mentioned earlier.

We now turn to the question of how to translate the results of finite dimensional model to the case of LSQM which is the content of the third question. Given that the theory is being expanded around a submanifold, such results are in general expressed in terms of various tubular expansion coefficients which are tensors of the ambient space evaluated on the submanifold. Since the Riemannian structure of $\cLM$ is induced from that of $\cM$, one would expect that all the relevant tubular expansion coefficients should be related to certain intrinsic geometric data of $\cM$. Finding such relations for the metric-expansion coefficients up to quadratic order will be the precise quantitative question addressed in \S \ref{s:tubular}. There are several technical steps to be followed in order to arrive at the final result which we explain in a self-contained manner in \S \ref{s:tubular}\footnote{In more technical terms, the final goal of \S \ref{s:tubular} is to develop a precise understanding of the metric-epansion given in (\ref{g-expansion}) in the context of loop space. This is done by suitably constructing (1) the tubular neighborhood of $\cM \hookrightarrow \cL \cM$ (content of \S \ref{ss:construct-tubular}) and (2) the FNC in $\cL \cM$ (content of \S \ref{ss:FNCinLM}). A construction of the relevant tubular neighborhood appeared before in \cite{stacey}. Although the general ideas are similar, our detailed construction is different and chosen to suit our purpose of constructing FNC. In particular, \cite{stacey} uses a method of embedding $\cM$ in a higher dimensional Euclidean space, wherease we use exponential maps. This enables us to relate the tubular geometry in $\cL \cM$ to a Riemannian structure in $T\cM$.}. Once these relations are known, one can use the precise analogy between the toy model and LSQM to translate results of \S \ref{s:finite-dim}. This will be discussed in \S \ref{s:analogy}. We conclude in \S \ref{s:conclusion} with some future directions. A brief note on loop space and LSQM that will be relevant for our discussion has been given in appendix \ref{a:note}. Appendix \ref{a:tubeH} and \ref{a:existence} contain some technical details.

\section{Tubular expansion of metric up to quadratic order}
\label{s:FNC}

Here we describe the basic set up for submanifold embedding that will be used throughout the paper. We consider a $D$-dimensional subspace $M$ embedded in a higher dimensional (pseudo) Riemannian space $L$ of dimension $d$. We adopt the following notations. Greek indices ($\alpha, \beta, \cdots$) run over $D$ dimensions, capital Latin indices ($A, B, \cdots$) run over $(d-D)$ dimensions and small Latin indices ($a, b, \cdots$), over all $d$ dimensions. The coordinates of $L$ will be denoted by $z^a=(x^{\alpha}, y^A)$ where $x^{\alpha}$ is a general coordinate system in $M$. Indices kept inside parenthesis will refer to non-coordinate basis, $\eta_{(ab)}$ being the diagonal matrix with the indictors as diagonal elements.

In \cite{florides} Florides and Synge (FS) proved existence of certain submanifold based coordinate system, called FNC in modern literature \cite{tubes}, which satisfies special coordinate conditions. In the special case where $M$ is a point, FNC reduces to RNC. The FS coordinate conditions can be described as follows. Equation for the submanifold is given by,
\bea
y^A = 0~, 
\eea 
The metric components of $L$, denoted by $g_{ab}(z)$, satisfy the following equations,
\bea
g_{a B}(z)y^B &=& \bar g_{a B}(x) y^B~, \cr
\bar g_{\alpha B} (x) &=& 0~, \cr
\bar g_{AB}(x) &=& \eta_{AB}~, 
\label{coord}
\eea
where $\bar g_{a b}(x) = g_{a b}(x, y=0)$. As a general rule, we use the lower case symbols to denote the geometric quantities of $L$ and the same symbols with bars to denote the same quantities restricted to the submanifold. With this convention in mind we will refrain from explicitly writing down the arguments of such quantities most of the time.  

The results for the expansion of the metric components away from the submanifold that will be relevant for us later are given by,
\bea
g_{\alpha \beta} &=& G_{\alpha \beta} + \bar s_{\alpha  \beta C} y^C  + (\bar \omega_{\alpha}{}^{\gamma}{}_C \bar \omega_{\beta \gamma D} + \bar \omega_{\alpha}{}^B{}_C \bar \omega_{\beta B D} + \bar r_{\alpha CD \beta}) y^C y^D + O (y^3)~, \cr
g_{\alpha B} &=& \bar \omega_{\alpha BC} y^C + {2\over 3} \bar r_{\alpha CDB} y^C y^D + O (y^3)~, \cr
g_{AB} &=& \eta_{AB} + {1\over 3} \bar r_{ACDB} y^C y^D + O (y^3) ~,
\label{g-expansion}
\eea
where $\omega{}_a{}^{(b)}{}_{(c)}$ are components of the connection one-form of $L$ (non-coordinate indices are converted to coordinate indices with the use of vielbein as usual),  $r_{acdb}$ is the covariant Riemann curvature tensor\footnote{We follow the same convention for the curvature as in \cite{nakahara}. } and
\bea
\bar s_{\alpha \beta C} = \bar \omega_{\alpha \beta C} + \bar \omega_{\beta \alpha C}~,
\label{second-fund}
\eea
is the second fundamental form of the submanifold embedding \cite{kobayashi-nomizu}.

We obtain (\ref{g-expansion}) from a closed form expression for the expansion of vielbein which is derived in \cite{tubular}. Although the details of this result will not be directly used in this work, there will be some relevance in the discussion of \S \ref{s:tubular}. We therefore summarize the main results of \cite{tubular} in appendix \ref{a:vielbein}.

\section{Finite dimensional analogue of loop space quantum mechanics}
\label{s:finite-dim}

In this section we will consider a finite dimensional analogue of LSQM in the framework discussed in \cite{dwv}. In \S \ref{ss:def} we will define the model and its semi-classical expansion. The analogue of linearized effective equation for tachyon fluctuation at leading order will be derived in \S \ref{ss:tachyon}. Our discussion below will be done without any reference to LSQM. We will come back to the analogy later in \S \ref{s:analogy}. 

\subsection{Definition of model and semi-classical expansion of Hamiltonian}
\label{ss:def}

All our notations used in the previous section will be valid in this section. We consider a non-relativistic quantum mechanical system whose configuration space is given by $L$. Hence it is assumed (only in this section) to have Euclidean signature. The Hamiltonian of the system is given by the standard expression,
\bea
\la \chi'|H^{pre}|\psi' \ra &=& \int dw \, \chi'^*(z) \cH^{pre} \psi'(z)~, \cr
\cH^{pre} &=& - {\hbar^2 \over 2} \cD^2 + V~,
\label{Hpre}
\eea
where $dw = dz \sqrt{g}$ is the invariant measure, $\cD^2$ is the Laplacian of $L$ and $V$ is a potential. We will consider $V$ to be confining to the submanifold $M \hookrightarrow L$ (a more precise definition will follow). We must define what we mean by performing a semi-classical expansion such that in the semi-classical limit the wavefunction collapses on the submanifold. This is a procedure given by the following steps\footnote{Various other, but similar, procedures have been discussed in the literature indicated earlier. Our procedure is adopted 
to suit LSQM.},
\begin{enumerate}
\item
{\bf Submanifold based description:} \\
Given $H^{pre}$ as in (\ref{Hpre}), we first move to a submanifold based description where the natural measure is given by $dy dx \sqrt{G}$ instead of $dw \sqrt{g}$\footnote{Recalling our notations introduced in \S \ref{s:FNC}, $dx \sqrt{G}$ is the invariant measure on the submanifold with respect to the induced metric.}. As discussed in appendix \ref{a:tubeH}, this is done by performing certain rescaling of the wavefunction so that the same matrix element in (\ref{Hpre}) is given in terms of the transformed Hamiltonian, which we call $H^{sub}$, and transformed wavefunctions (unprimed) as,
\bea
\la \chi|H^{sub}|\psi \ra &=& \int dy dx \sqrt{G} \, \chi^*(z) \cH^{sub} \psi(z)~,
\label{cHsub}
\eea
where the expression for $\cH^{sub}$ can be found in eq.(\ref{cHsub-result}). 

\item
{\bf Tubular expansion:}\\
Next we tubular expand $\cH^{sub}$ to write,
\bea
\la \chi|H^{sub}|\psi \ra &=& \sum_{n=0}^{\infty} \int dy dx \sqrt{G} \, \chi^*(z) \cH^{sub}_n \psi(z)~, 
\label{cHsub-n}
\eea
where we adopt the following notation: $X_n$ is the contribution at $O(y^n)$ in the tubular expansion of $X$.
Explicit result for $\cH^{sub}_n$ are given in eqs.(\ref{cHsubn-result}, \ref{Kn-result}, \ref{Dn-dn-tn-result}).

\item 
{\bf Definition of $V$:}\\
The potential $V$ is confining to the submanifold $M$ whose embedding satisfies the following property,
\bea
\bar \omega_{\alpha \beta C} = 0~.
\label{vanishing-omega}
\eea
As a result the second fundamental form in (\ref{second-fund}) vanishes and therefore $M$ is totally geodesic \cite{kobayashi-nomizu}. Furthermore, the transverse profile of $V$ is given by the following expression in FNC\footnote{We will display the summation over indices explicitly, as we have done in eq.(\ref{pot}), whenever Einstein summation convention will not be valid.},
\bea
V(x, y) &=& {1\over 2} \sum_{A, B} \eps_A \eps_B g_{A B} y^A y^B~,   
\label{pot}
\eea
where $\eps_A$ is positive definite\footnote{Notice that due to the presence of the $\eps$-factors in the potential, general covariance of the ambient manifold is broken down to that of the submanifold even at the classical level. Without such factors $V$ will be a scalar under the full diffeomorphism of $L$, but will not have a non-trivial tubular expansion due to the coordinate condition in (\ref{coord}). As will be explained in \S \ref{s:analogy}, general covariance gets broken in LSQM only due to the semi-classical limit. }.

\item
{\bf Definition of semi-classical expansion:} \\
Define a rescaled Hamiltonian,
\bea
H = {1\over \hbar} H^{sub} ~.
\label{H-def}
\eea
Semi-classical expansion of $H$ is given by rescaling the transverse coordinates as, 
\bea
y^A \to \sqrt{\hbar \over \eps_A} y^A~.
\label{y-rescaling}
\eea
in the tubular expansion of $H$. This gives,
\bea
\la \chi|H|\psi \ra &=& \sum_{n=0}^{\infty} \hbar^{n\over 2} \int dy dx \sqrt{G}  \chi^*(z) {}^{\eps}\cH^{(n)} \psi(z)~, 
\label{hbar-exp-H2}
\eea
where we have used the following notation,
\bea
{}^{\eps}X &=& X|_{y^A \to {y^A/ \sqrt{\eps_A}}}~.
\label{eps-notation}
\eea
\end{enumerate} 

The expression for ${}^{\eps}\cH^{(n)}$ can be found in eq.(\ref{calH-n2}). For the rest of our analysis we will restrict the expansion in (\ref{hbar-exp-H2}) up to $O(\hbar)$. Explicit computation using the metric-expansion in (\ref{g-expansion}) yields the following results,
\bea
{}^{\eps}\cH^{(0)} &=& {1\over 2} \sum_{A, B} \sqrt{\eps_A \eps_B} (-\eta^{A B} \del_A \del_B + \eta_{A B} y^A y^B) 
= \sum_{A, B} \sqrt{\eps_A \eps_B} \eta_{A B} a^{\dagger}{}^A a^B + {1\over 2} \sum_A \eps_A ~, \cr
{}^{\eps}\cH^{(1)} &=& 0~, \cr
{}^{\eps}\cH^{(2)} &=& - {1\over 2} (\nabla^{\alpha} + i \bar \omega^{\alpha A B} \,\, {}^{\eps}\Lambda_{A B}) (\nabla_{\alpha} + i \bar \omega_{\alpha}{}^{C D} \,\, {}^{\eps}\Lambda_{C D})  - {1\over 4} \bar r_{\parallel} - {1 \over 12} \bar r_{\perp}  \cr
&& + {1\over 6} \bar r^{A B C D} \,\, {}^{\eps}\Lambda_{A B} \,\, {}^{\eps}\Lambda_{C D} 
+ {1 \over 6} \sum_{A, B} \sqrt{\eps_A \eps_B \over \eps_C \eps_D} \, \bar r_{A C D B} y^A y^C y^D y^B ~, 
\label{epsH012}
\eea
where in the first equation we have defined annihilation and creation operators,
\bea
a^A = {1\over \sqrt{2}}(\eta^{A B} \del_B + y^A) ~, \quad a^{\dagger}{}^A = {1\over \sqrt{2}}(-\eta^{A B} \del_B + y^A) ~,
\eea
respectively, such that $[a^A, a^{\dagger}{}^B] =  \eta^{AB}$. In the last equation $\nabla_{\alpha}$ denotes the covariant derivative with respect to the induced metric $G_{\alpha \beta}(x)$ on $M$. The other new notations introduced in these equations are as follows,
\bea
\bar r_{\parallel} &=& \bar r^{\alpha B}{}_{\alpha B}~, \quad \bar r_{\perp} = \bar r^{A B}{}_{A B}~,\cr
{}^{\eps}\Lambda_{A B} &=& -{i \over 2} (\sqrt{\eps_B \over \eps_A} \, \eta_{A C} y^C \del_B 
- \sqrt{\eps_A \over \eps_B} \, \eta_{B C} y^C \del_A)~.
\eea
As noted in the literature, $\Lambda_{A B}$ is the angular momentum operator in the transverse space and $\bar \omega_{\alpha A B}$ is analogous to a non-abelian ($SO(d-D)$) Berry connection \cite{berry}. As a mathematical exercise, our results in eqs.(\ref{epsH012}) are equivalent to a case discussed in \cite{mitchell} except for the last term in the last equation which comes from the tubular expansion of our potential.

The reasons why the above procedure correctly captures our general idea of Born-Oppenheimer type approximation and localization of wavefunction are as follows. The $\hbar$ dependent rescaling in eq.(\ref{y-rescaling}) makes the transverse coordinates $y$ {\it fast} in the Born-Oppenheimer sense. As a result, the leading order harmonic oscillator Hamiltonian, i.e. ${}^{\eps}\cH^{(0)}$ is independent of the {\it slow} coordinates ($x$). Therefore, the transverse and longitudinal dynamics decouple. Moreover, ${}^{\eps}\cH^{(0)}$ is $\hbar$-independent. The wavefunctions fall of to zero at large values of the rescaled coordinates. At leading order this corresponds to arbitrary finite values of the original transverse coordinates indicating that the wavefunctions are localized.

\subsection{Analogue of linearized tachyon effective equation at leading order}
\label{ss:tachyon}

Here we consider the transverse degrees of freedom to be frozen in the harmonic oscillator ground state and derive the 
{\it effective Hamiltonian}, as will be defined in eq.(\ref{H-Heff}) below, for the longitudinal degree of freedom at the leading order. This will give us the linearized tachyon effective equation (see eq.(\ref{tachyonEOM})) at this order. We will explain this analogy later in \S \ref{s:analogy}.

The wavefunction under consideration is,
\bea
\psi_0(x, y) &=& T(x) \chi_0(y)~,
\eea
such that,
\bea
a^A \chi_0(y) &=& 0~,  \quad \int dy ~\chi_0^2(y) = 1~.
\eea
The expectation value of $H$ up to first order in $\hbar$ is gievn by,
\bea
\la \psi_0|H|\psi_0 \ra &=& \int dx \sqrt{G} \, T^*(x) \la ({}^{\eps}\cH^{(0)} + \hbar {}^{\eps}\cH^{(2)}) \ra^{\perp}_0T (x)~, 
\label{hbar-exp-H3}
\eea
where we have used,
\bea
\la {\cal O} \ra^{\perp}_0 &=& \int dy ~\chi_0(y) {\cal O} \chi_0(y)~.
\eea
Using the results,
\bea
\la {}^{\eps}\cH^{(0)} \ra^{\perp}_0 &=& {1\over 2} \sum_A \eps_A~, \cr
\la {}^{\eps}\Lambda_{A B} \ra^{\perp}_0 &=& 0~, \cr
\la {}^{\eps}\Lambda_{A B} {}^{\eps}\Lambda_{C D} \ra^{\perp}_0 &=& {(\eps_A - \eps_B)^2 \over 16 \eps_A \eps_B} (\eta_{A C} \eta_{B D} - \eta_{A D} \eta_{B C}) ~, \cr
\la y^A y^C y^D y^B \ra^{\perp}_0 &=& {1\over 4} (\eta^{C D}\eta^{A B} + \eta^{C B} \eta^{A D} + \eta^{A C} \eta^{D B})~,
\label{VEVs}
\eea
one finds for the effective Hamiltonian $\cH_{eff}$ for $T(x)$, 
\bea
\la \psi_0|H|\psi_0 \ra &=& \hbar \int dx \sqrt{G(x)} \, T^*(x) \cH_{eff} T(x)~, \cr
\cH_{eff} &=& -{1\over 2} \nabla^2 + {m^2 \over 2} + V_{eff}^{\bar r}(x) + V_{eff}^{\bar \omega}(x) ~,
\label{H-Heff}
\eea
where,
\bea
m^2 &=& {1\over \hbar} \sum_A \eps_A ~, \cr
V_{eff}^{\bar r}(x) &=& -{1\over 4} \bar r_{\parallel} - {1\over 12} \bar r_{\perp} + \sum_{A, B} \lt[{(\eps_A - \eps_B)^2\over 48 \eps_A \eps_B} - {(\eps_A -\eps_B)\over 24 \eps_B} \rt] \bar r^{A B}{}_{A B} ~, \cr
V_{eff}^{\bar \omega}(x) &=& \sum_{B, D} {(\eps_B - \eps_D)^2 \over 16 \eps_B \eps_D} \bar \omega^{\alpha B D} \bar \omega_{\alpha B D} ~.
\label{Heff-comp}
\eea
As the reason will be explained in \S \ref{s:analogy}, we identify the following equation as the analogue of linearized tachyon effective equation,
\bea
\cH_{eff} T(x) = 0~.
\label{tachyonEOM}
\eea 

\section{Tubular neighborhood of target manifold in loop space}
\label{s:tubular}

The leading order analysis of the finite dimensional model as done in the previous section will be interpreted in the context of LSQM in \S \ref{s:analogy}. As motivated in \S \ref{s:intro}, the submanifold of interest in this case is the space of all constant loops which is isomorphic to $\cM$ itself (see appendix \ref{a:note}). The main ingredient that has gone into the analysis of \S \ref{s:finite-dim} is the tubular expansion of the ambient space metric up to quadratic order as given in eqs.(\ref{g-expansion}) with the additional condition (\ref{vanishing-omega}). The goal of this section will be to understand the analogue of this expansion in $\cLM$. More precisely, we will show that eqs.(\ref{g-expansion}, \ref{vanishing-omega}) will still be valid, with the notations correctly interpreted, for the embedding $\cM \hookrightarrow \cLM$ with suitably constructed FNC and with additional equations, given in (\ref{tub-coeff-LM}), that relate the relevant expansion coefficients to intrinsic geometric data of $\cM$. 

The steps that we will follow are as follows. First we present an explicit construction of the tubular neighborhood in \S \ref{ss:construct-tubular}. A proof of existence was given through an explicit construction earlier by Stacey in \cite{stacey}. Although the basic ideas are similar, the details of our construction are different and has been chosen to suit our purpose (in \S \ref{ss:FNCinLM}) better. The result of \S \ref{ss:construct-tubular} will show how the relevant Riemannian structure in $\cLM$ is related to a certain Riemannian structure on the tangent bundle $T \cM$ of $\cM$. The latter views the zero-section $T\cM_0$ (which is isomorphic to $\cM$) as a submanifold embedded in $T \cM$ such that the normal bundle $N(T\cM_0)$ is isomorphic to $T \cM$ itself. 
In \S \ref{ss:FNCinLM} we first compute the tubular expansion of the metric in $T \cM$ up to quadratic order using a procedure that is implicit in the construction of \S \ref{ss:construct-tubular}. The expansion coefficients are all related to intrinsic geometric data of $\cM$.\footnote{For any given Riemannian structure on $T \cM$, one would expect that the geometric data of $T \cM$ should be expressible in terms of those of $\cM$. See for example \cite{sasaki}.} Such relations have been found up to an undetermined real parameter which is not fixed at the present level of approximation by the analogue of eq.(\ref{diff-viel}) which determined the tubular expansion of vielbein in the finite dimensional context. Finally, we derive the metric expansion in $\cLM$ (up to quadratic order) using the aforementioned relation between $\cLM$ and $T \cM$. Therefore the ambiguity of the real parameter mentioned above is carried over to $\cLM$. However, as explained in \S \ref{s:analogy}, this ambiguity does not affect our conclusion about the tachyon effective equation. 

\subsection{Explicit construction of the tubular neighborhood}
\label{ss:construct-tubular}

Below we will first heuristically describe our construction and then specify it in more mathematical terms, in particular make connection with \cite{stacey}. We will refer to geodesics and open neighborhoods in both $\cLM$ and $\cM$ in various places along the way. It should be clear from the context which space we are referring to.

\begin{wrapfigure}{l}{5cm} 
\centering
\includegraphics[height=35mm]{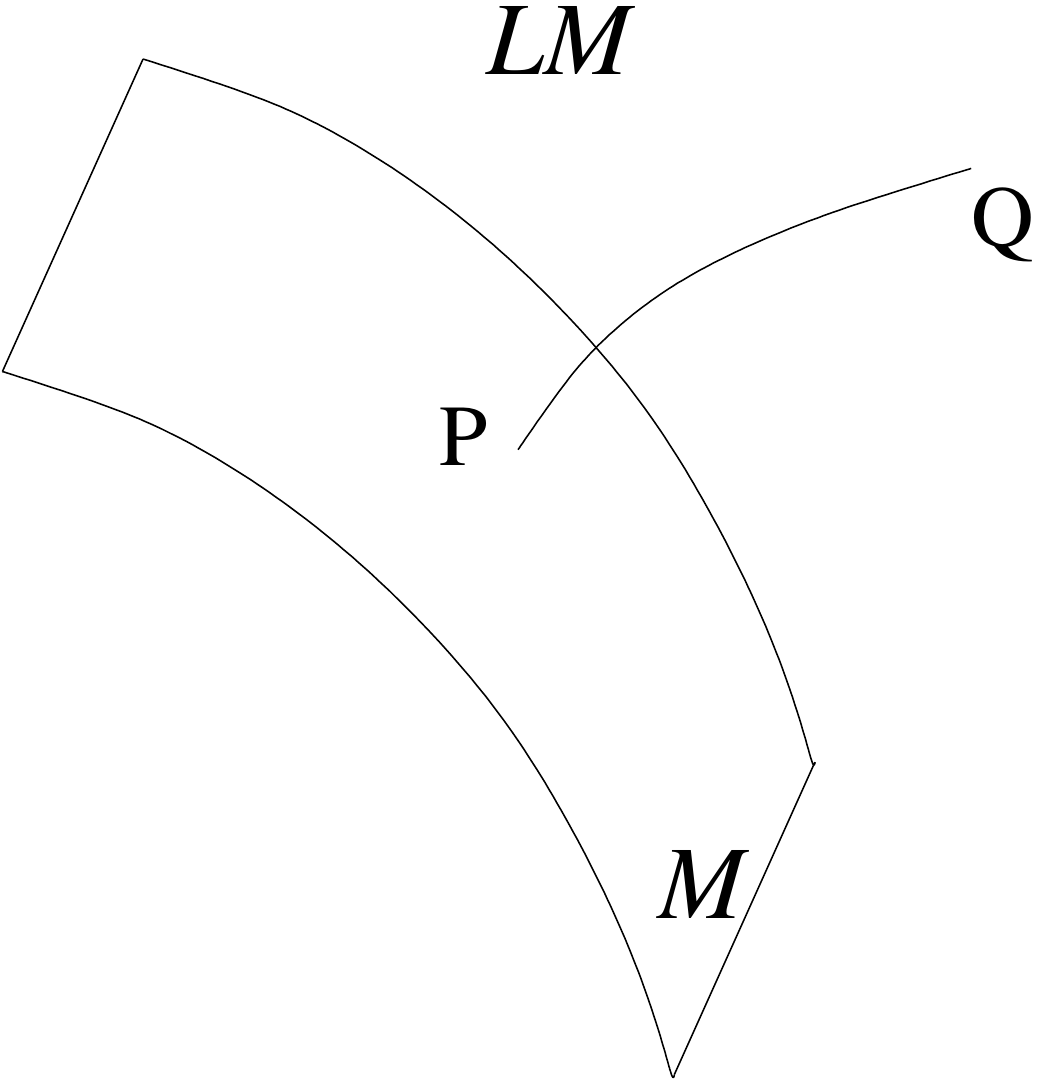}
\caption{Tubular neighborhood of $\cM$ in $\cLM$.}
\label{f:tubnbhd}
\end{wrapfigure}

The basic picture \cite{florides}, true for any tubular neighborhood, that we will have in mind is given in fig.\ref{f:tubnbhd}. Given any point $Q$ in the neighborhood, there exists a unique geodesic passing through $Q$ that arrives at a unique point $P$ on the submanifold orthogonally. 
This condition is not satisfied if two geodesics emerging orthogonally from the submanifold meet at a point outside. Following the standard way, we will restrict ourselves to a  region sufficiently close to the submanifold such that this does not happen. Recall that every point in the neighborhood in $\cLM$ corresponds to a non-zero loop in $\cM$ such that nearer the point resides to the submanifold of vanishing loops, smaller the loop it represents. It turns out that the above restriction corresponds to considering sufficiently {\it small loops} in $\cM$ such that any given loop can be entirely encompassed within a single convex normal neighborhood \cite{wald84} in $\cM$. This implies that a small loop should fit entirely into $B_{p}$ - the ball of largest RNC-radius with center at $p \in M$, for some $p$ in the neighborhood.

\begin{wrapfigure}{l}{5cm} 
\centering
\includegraphics[height=35mm]{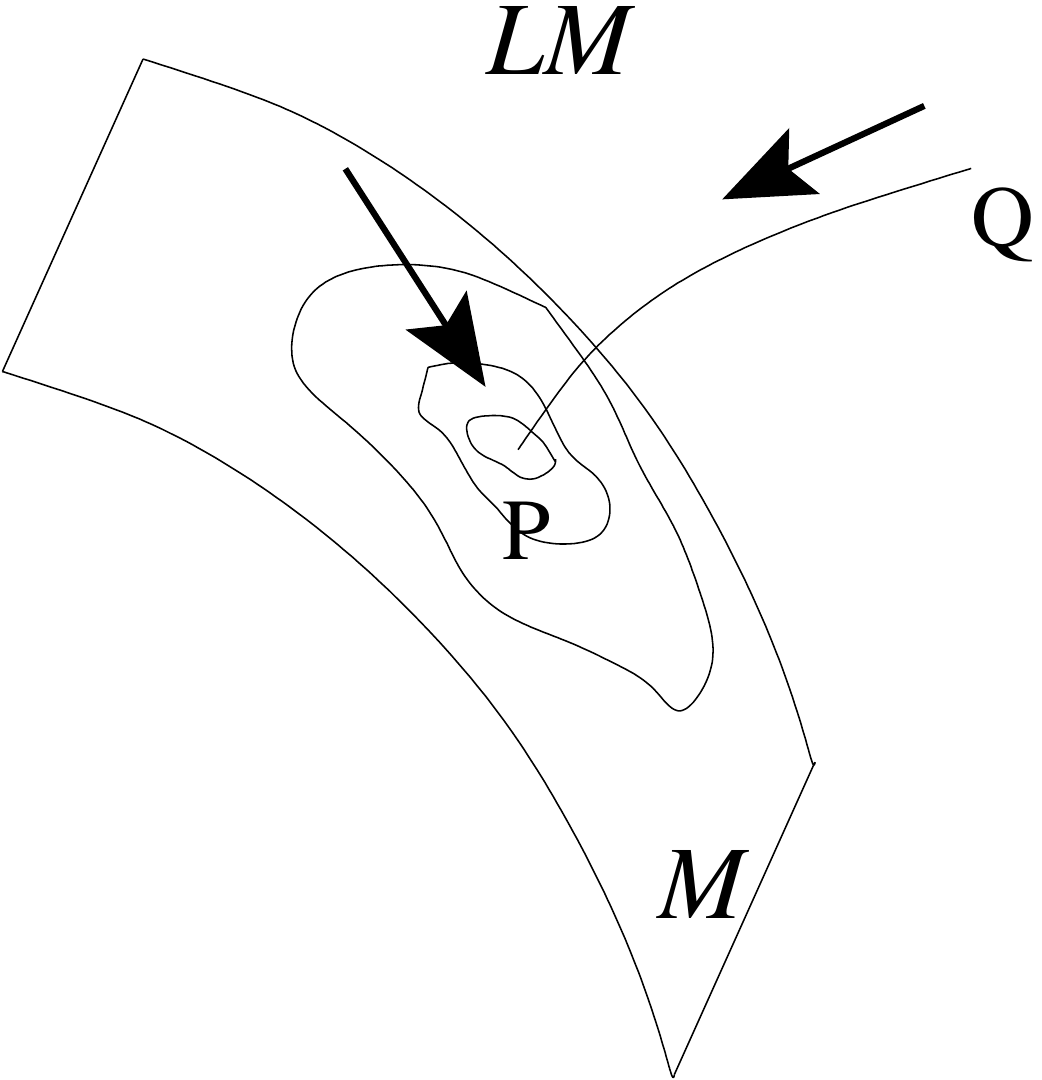}
\caption{Average position of loops.}
\label{f:loop-ave}
\end{wrapfigure}

Let us now consider the set of points lying on the geodesic $QP$ in $\cLM$. This corresponds to a class of loops which progressively shrink to zero size as we approach $P$ (see cartoon in fig.\ref{f:loop-ave}). Therefore from the perspective of the interior of $\cM$ this defines $P$ to be some kind of an {\it average value} for all the loops in this class. Notice that such a definition of averaging is independent of the choice of coordinate system, simply because it only refers to geodesics.
Given a loop-embedding $Z$ in $\cM$, its average position, as defined above, can be found in the following way. In case $\cM$ is flat, it is simply given by $\int_{S^1} Z$. This is basically because Minkowski space is also a vector space where one can define a radial vector. In a curved space one should make use of geodesics which look like radial vectors in RNC, the latter being related to general coordinates through the exponential map. Therefore, when $\cM$ is curved, we first describe the loop in RNC centered at a suitable point, with coordinate say $\underline{x}$. The choice of this base-point is not fixed, as the loop will in general fit into $B_{\underline{x}}$ for a range of values of $\underline{x}$. However, there is a unique value $x$ within the allowed range for which the following condition is satisfied,
\bea
\int_{S^1} \hat Y = 0~,
\label{loop-ave-TxM}
\eea
where $\hat Y$ is the loop in RNC centered at $x$. Given the initial loop in $\cM$, we identify $x$ as the average position. This way every small loop is uniquely assigned an average position in $\cM$. Moreover, every point in a suitable neighborhood of $\cM$ is uniquely assigned to a class of small loops through the above procedure which is viewed to form the directions nomal to $\cM$ in $\cLM$. 

To facilitate the discussion in next subsection we now describe the above construction in more technical terms. We begin by introducing the following notation: $U_S$ is open in $S$ where $S$ stands for $\cM$, $\cM \times \cM$, $T \cM$ and $\cLM$. We will see how these various open neighborhoods are inter-related. Given an open normal neighborhood $U_{\cM} \subset \cM$, let $C(U_{\cM})$ be the collection of all small loops such that the average positions of all such loops are contained in $U_{\cM}$. We also assume that all the loops in $C(U_{\cM})$ are contained in a single open normal neighborhood $\tilde U_{\cM} \supset U_{\cM}$. Each element $l \in C(U_{\cM})$ can be associated to an element $(x, Z(\s)) \in U_{\cM \times \cM}$ where $Z(\s)$ is the embedding of the loop $l$ given in general coordinates (see appendix \ref{a:note}), $x$ is the average position of the loop and $U_{\cM \times \cM}$ is a suitable open neighborhood containing all the loops in $C(U_{\cM})$. The pre-image of the loop under the exponential map found in $T_x \cM$ is given by $\hat Y(\s)$ which satisfies eq.(\ref{loop-ave-TxM}). Repeating this procedure for all the loops in $C(U_{\cM})$ one arrives at the following set \cite{stacey}, 
\bea
U_{\cLM} &=& \lt\{ \hat Y : S^1 \to U_{T\cM} , \pi \hat Y \hbox{ is constant }, \int_{S^1} \hat Y = 0 \rt\}~,
\label{stacey-eqn}
\eea
where $\pi: U_{T\cM} \to U_{\cM}$ is the projection map. Constancy of $\pi \hat Y$ implies that the whole loop resides in the same fibre, unlike its configuration in $\cM \times \cM$. 

Since exponential map is a diffeomorphism, the above argument shows that the desired tubular neighborhood is diffeomorphic to the set in (\ref{stacey-eqn}). The relevant diffeomorphism is a bundle map which is the collection of all the inverse exponential maps at all $x\in U_{\cM}$,
\bea
\exp^{-1}: U_{\cM \times \cM} \to U_{T \cM}~.
\label{bundle-map}
\eea
If $\Dlt$ is the diagonal submanifold of $\cM \times \cM$, then $\exp^{-1}$ maps $U_{\cM \times \cM} \cap \Dlt$ to $U_{T \cM} \cap T \cM_0$, both being isomorphic to $U_{\cM}$. This kind of a construction is called a {\it local addition} (see \cite{KM} for a precise definition), of which the exponential map is a standard example. In \cite{stacey}
construction of the local addition has been facilitated by embedding $\cM$ in a higher dimensional Euclidean space. However, exponential map is more suitable for our purpose, as we will see in the next subsection where the aforementioned diffeomorphism will be explicitly constructed. 

\subsection{Construction of FNC}
\label{ss:FNCinLM}

Here we would like to understand the loop space analogue of the metric expansion given in (\ref{g-expansion}). As mentioned before, generically the tubular expansion coefficients are certain geometric quantities of the ambient manifold evaluated on the submanifold. Although these are not related to the intrinsic geometric properties of the submanifold in general, for loop space that is the case. Therefore, expressing the tubular expansion coefficients of $\cLM$ in terms of the geometric data of $\cM$ is the precise quantitative question that needs to be answered. The discussion in the previous subsection implicitly defines a procedure to answer this question which we pursue here. 

There are two steps to be followed. Given a Riemannian structure on $\cM$, the space $\cM \times \cM$ acquires a natural direct product structure. The bundle map (diffeomorphism) in (\ref{bundle-map}) enables one to view $T \cM$ as a Riemannian manifold where $T\cM_0$ sits as a submanifold whose normal bundle $N(T \cM_0)$ is isomorphic to $T \cM$. The first step is to construct the relevant submanifold based coordinate system, which we call $\hbox{FNC}_{U_{T \cM}}$, on $U_{T \cM}$ by a suitable coordinate transformation from the direct product coordinate system on $U_{\cM \times \cM}$. This will be discussed in \S \ref{sss:TM}. Then the final step is to construct FNC in $\cLM$ by {\it looping} $\hbox{FNC}_{U_{T \cM}}$, a procedure that has been explained in \S \ref{sss:looping}.

\subsubsection{Riemannian structure on $T \cM$}
\label{sss:TM}

In \S \ref{s:FNC} we considered a submanifold $(M, G)$ embedded in an ambient space $(L, g)$ such that $G$ is the induced metric obtained from $g$. Here we consider a special case where $(T \cM_0\cong \cM , G)$ is embedded in $(T \cM, \hat g)$. The speciality of this case is that the tubular expansion coefficients are related to quantities obtained from the basic data $(\cM, G)$ \footnote{Because of the involvement of spin connection, we will see that the basic data is actually given by the vielbein of $\cM$.}. Below we will construct $\hbox{FNC}_{U_{T \cM}}$ up to quadratic order by starting with $(\cM\times \cM, \bar g)$ and performing suitable coordinate transformations. There will be a certain degree of indeterminacy in our final result which, as will be explained toward the end of this subsection, is not resolved at the present level of approximation by the analogue of eq.(\ref{diff-viel}).

We begin by discussing $(\cM\times \cM, \bar g)$. The coordinates of a point in $U_{\cM \times \cM}$ are given by,
\bea
\bar z^a = (x^{\alpha_1}_1, x^{\alpha_2}_2)~, \quad \alpha_1, \alpha_2 = 1, \cdots, D (=\dim \cM) ~.
\label{z-bar}
\eea
The components of the vielbein are given by,
\bea
\bar e^{[\alpha_1]}{}_b (\bar z) &=& \pmatrix{E^{(\alpha_1)}{}_{\beta_1}(x_1) \cr 0 }~, \quad 
\bar e^{[\alpha_2]}{}_b (\bar z) = \pmatrix{0 \cr E^{(\alpha_2)}{}_{\beta_2}(x_2)}~,
\label{ebar-comp}
\eea
Indices in square brackets refer to the non-coordinate basis in $\cM \times \cM$. $E^{(\alpha)}{}_{\beta}$ are the vielbein components of $\cM$ (with indices in round brackets referring to the non-coordinate basis) with metric components given by $G_{\alpha \beta}$,
\bea
G_{\alpha \beta} = E^{(\gamma)}{}_{\alpha}E_{(\gamma) \beta}~.
\label{M-metric}
\eea
In general the two copies of $\cM$ can have different metrics which are diffeomorphic to each other. We have chosen $x_1$ and $x_2$ suitably so that these two metrics are same as given in eq.(\ref{M-metric}). We will denote the desired coordinate system $\hbox{FNC}_{U_{T \cM}}$ by $\hat z = (x^{\alpha},\hat y^{\hat \alpha})$ with $\alpha, \hat \alpha = 1, \cdots , D$\footnote{Notice that the indices $\alpha_1$, $\alpha_2$, $\alpha$ and $\hat \alpha$ all run over $D= \dim \cM$ dimensions. From the perspective of $\cM$ they do not make any difference. However, from the perspective of $\cM \times \cM$ they do.}, which will be obtained below by following a series of coordinate transformations from $\bar z$. 

The first step is to argue, as has been done in appendix \ref{a:existence}, that there exists a coordinate system, 
\bea
z'^a = (x'^{\alpha}, y'^{\hat \alpha})~,
\label{z'}
\eea 
where $x'$ is a general coordinate system on $\Dlt_{U_{\cM \times \cM}} = U_{\cM \times \cM} \cap \Dlt$ such that the transformed components of the vielbein (with an additional overall constant scaling of the metric, see appendix \ref{a:existence}) are given by the following expansion up to quadratic order in $y'$.
\bea
e'^{[\alpha]}{}_{\beta} (z') &=& E^{(\alpha)}{}_{\beta}(x') + q \check R^{(\alpha)}{}_{\hat \gamma \hat \delta \beta}(x') y'^{\hat \gamma} y'^{\hat \delta}~, \cr
e'^{[\alpha]}{}_{\hat \beta} (z') &=& 0~, \cr
e'^{[\hat \alpha]}{}_{\beta} (z') &=& 0~, \cr
e'^{[\hat \alpha]}{}_{\hat \beta} (z') &=& E^{(\hat \alpha)}{}_{\hat \beta}(x') + {1\over 6} \check R^{(\hat \alpha)}{}_{\hat \gamma \hat \delta \hat \beta}(x')  y'^{\hat \gamma} y'^{\hat \delta}~,
\label{arbit-prime}
\eea 
where the orthonormal frames with superscripts $[\alpha]$ and $[\hat \alpha]$ are parallel and transverse to the submanifold respectively and are obtained from the ones in eqs.(\ref{ebar-comp}) through a rotation of the local Lorentz frame as given in eqs.(\ref{local-rotation}). The symbol $\check R^{(\alpha)}{}_{\gamma \delta \beta}(x')$ denotes 
the Riemann curvature tensor component $R^{(\alpha)}{}_{\gamma \delta \beta}$ \footnote{We use upper case symbols without a $\check{}$ to denote tensors of $\cM$ in general coordinates.} of $\cM$ evaluated at $x'$ in $\hbox{RNC}_{x'}$ where $\hbox{RNC}_{x'}$ refers to the RNC-system centered at $x'$ such that the vielbein components are given by $E^{(\alpha)}{}_{\beta}(x')$ at the centre. $q$ is an undetermined real number. We will argue toward the end of this subsection that the analogue of eq.(\ref{diff-viel}) is satisfied up to quadratic order for arbitrary values of $q$. 

Then the final step is to perform the following coordinate transformation: 
$z'^a \to \hat z^a = (x^{\alpha}, \hat y^{\hat \alpha})$ such that,
\bea
x^{\alpha} = x'^{\alpha}~, \quad \hat y^{\hat \alpha} = E^{(\hat \alpha)}{}_{\hat \beta}(x) y'^{\hat \beta}~.
\eea
The transformed vielbein components are given by,
\bea
\hat e^{[\alpha]}{}_{\beta}(\hat z) &=& E^{(\alpha)}{}_{\beta}(x) + q \check R^{(\alpha)}{}_{\hat \gamma \hat \delta \beta}(x) \hat y^{\hat \gamma} \hat y^{\hat \delta}~, \cr
\hat e^{[\alpha]}{}_{\hat \beta}(\hat z) &=& 0~, \cr
\hat e^{[\hat \alpha]}{}_{\beta}(\hat z) &=& E^{(\hat \alpha)}{}_{\hat \delta}(x) \del_{\beta} E_{(\hat \gamma)}{}^{\hat \delta}(x) \hat y^{\hat \gamma}~, \cr
\hat e^{[\hat \alpha]}{}_{\hat \beta}(\hat z) &=& \dlt^{\hat \alpha}{}_{\hat \beta} + {1\over 6} \check R^{(\hat \alpha)}{}_{(\hat \gamma \hat \delta \hat \beta)}(x) \hat y^{\hat \gamma} \hat y^{\hat \delta}~,
\label{vielbeinhat-expansion}
\eea
which give the following results for the metric components,
\bea
\hat g_{\alpha \beta} &=& G_{\alpha \beta} + \bar{\hat s}_{\alpha  \beta \hat \gamma} \hat y^{\hat \gamma}  
+ (\bar{\hat \omega}_{\alpha}{}^{\eta}{}_{\hat \gamma} \bar{\hat \omega}_{\beta \eta \hat \delta} 
+ \bar{\hat \omega}_{\alpha}{}^{\hat \eta}{}_{\hat \gamma} \bar{\hat \omega}_{\beta \hat \eta \hat \delta} 
+ \bar{\hat r}_{\alpha \hat \gamma \hat \delta \beta}) \hat y^{\hat \gamma} \hat y^{\hat \delta} + O(\hat y^3)~, \cr
\hat g_{\alpha \hat \beta} &=& \bar{\hat \omega}_{\alpha \hat \beta \hat \gamma} \hat y^{\hat \gamma} 
+ {2\over 3} \bar{\hat r}_{\alpha \hat \gamma \hat \delta \hat \beta} \hat y^{\hat \gamma} \hat y^{\hat \delta} 
+ O(\hat y^3)~, \cr
\hat g_{\hat \alpha \hat \beta} &=& \eta_{\hat \alpha \hat \beta} + {1\over 3} \bar{\hat r}_{\hat \alpha \hat \gamma \hat \delta \hat \beta} \hat y^{\hat \gamma} \hat y^{\hat \delta} + O(\hat y^3) ~,
\label{ghat-expansion}
\eea
with,
\bea
\bar{\hat \omega}_{\alpha \beta \hat \gamma} &=& 0~, \quad (\Rightarrow  \bar{\hat s}_{\alpha \beta \hat \gamma} = 0)~, \cr
\bar{\hat \omega}_{\alpha \hat \beta \hat \gamma} &=& E_{(\hat \beta) \hat \delta}(x) \del_{\alpha}E_{(\hat \gamma)}{}^{\hat \delta}(x) ~, \cr
\bar{\hat r}_{\alpha \hat \gamma \hat \delta \beta} &=& 2 q \check R_{\alpha (\hat \gamma \hat \delta) \beta}(x)~, \cr
\bar{\hat r}_{\alpha \hat \gamma \hat \delta \hat \beta} &=& 0~, \cr
\bar{\hat r}_{\hat \alpha \hat \gamma \hat \delta \hat \beta} &=& \check R_{(\hat \alpha \hat \gamma \hat \delta \hat \beta)}(x)~.
\label{TM-coeff}
\eea
A few remarks about eqs.(\ref{vielbeinhat-expansion}, \ref{ghat-expansion}) and (\ref{TM-coeff}) are in order. All the hatted variables appearing in these equations are tensors of $\cM \times \cM$ in $\hat z = (x, \hat y)$ coordinate system. Equivalently, they can also be viewed as tensors of $T \cM$. In particular,  eqs.(\ref{vielbeinhat-expansion}) and (\ref{ghat-expansion}) are viewed to describe the tubular expansion of vielbein and metric components respectively up to quadratic order in FNC. Notice that eqs.(\ref{ghat-expansion}) are written in the general form of tubular expansion as in (\ref{g-expansion}) following the same rules for notation adopted there. Therefore a bar indicates that a tensor of $T \cM$ is being evaluated on $U_{T \cM} \cap T \cM_0$. Equations (\ref{TM-coeff}) exhibit how such quantities are related to intrinsic geometric quantities of $\cM$. Finally, the results in (\ref{ghat-expansion}) satisfy the analogue of the coordinate conditions in (\ref{coord}) because of the anti-symmetry properties of the spin connection and curvature tensor identified in eqs.(\ref{TM-coeff})\footnote{Notice that the spin connection identified in the second equation of (\ref{TM-coeff}) is indeed anti-symmetric in the last two indices as required.}. 

The tubular expansion in (\ref{vielbeinhat-expansion}) is supposed to satisfy the analogue of the differential equation (\ref{diff-viel}) in $T\cM$. This is given, in our notation adopted here, by,
\bea
\hat{\hbd} (\hat{\hbd} + \mathop{\eps}^b) \hat e^{[a]}{}_b(\hat z) = \hat \rho^{[a]}{}_{[c]}(\hat z; \hat y) \hat e^{[c]}{}_b(\hat z)~,
\label{diff-viel-TM}
\eea
where $\hat{\hbd} = \hat y^{\hat \alpha} {\del \over \del \hat y^{\hat \alpha}}$, $a=\alpha, \hat \alpha$, $\mathop{\eps}^b$ being $1$ for $b=\hat \beta$ and $-1$, otherwise and,
\bea
\hat \rho^{[a]}{}_{[b]} (\hat z; \hat y') = \hat r^{[a]}{}_{\hat \gamma \hat \delta [b]}(\hat z) \hat y'^{\hat \gamma}
\hat y'^{\hat \delta}~.
\eea
It is straightforward to to check that eq.(\ref{diff-viel-TM}) is satisfied by eqs.(\ref{vielbeinhat-expansion}) up to quadratic order for arbitrary values of $q$ provided the curvature components of $T\cM$ are identified according to the last three equations in (\ref{TM-coeff}).

\subsubsection{Looping $\hbox{FNC}_{T\cM}$}
\label{sss:looping}

The desired FNC in $\cLM$ is obtained by looping the coordinate system $\hat z^{\hat a} = (x^{\alpha}, \hat y^{\hat \alpha})$ constructed in the previous subsection. To explain the method we first recall the following facts:
\begin{enumerate}
\item
The normal bundle $N(T \cM_0\cong M)$ is isomorphic to $T \cM$. $\hat y^{\hat \alpha}$ are the coordinates on the fibres $N_x(T \cM_0)$. 
\item
The desired tubular neighborhood in $\cLM$ is given by the space of non-zero loops in (\ref{stacey-eqn}). Here every loop resides entirely in a single fibre $T_x(\cM)$ such that the average of the loop is the corresponding base point $x$. 
\end{enumerate}

Therefore the general coordinates $x^{\alpha}$ on the submanifold $\cM \hookrightarrow \cLM$ is the same as that on $T \cM_0\hookrightarrow T \cM$. And the normal coordinates in the neighborhood of $\cM \hookrightarrow \cLM$ is given by the Fourier transforms (see appendix \ref{a:note}) of the loop in $T_x(\cM)$ as described above. In terms of equations, the FNC in $\cLM$ is given by,
\bea
z^a &=& (x^{\alpha}, y^A)~,
\label{FNC-LM}
\eea
where,
\bea
y^A &=& \oint \hat Y^{\hat \alpha}(\s) e^{-i \ha \s}~, \quad \ha \neq 0~,
\label{yA-def1}
\eea
such that $\hat Y^{\hat \alpha}(\s)$ satisfies eq.(\ref{loop-ave-TxM}).

Tensors in $\cLM$ are obtained from those in $T \cM$ following a similar procedure \cite{dwv} which we discuss in detail now. Let $\{\hat e_{\hat a}\} = \{\hat e_{\alpha}={\del \over \del x^{\alpha}}, \hat e_{\hat \alpha}={\del \over \del \hat y^{\hat \alpha}} \}$ and $\{d\hat z^{\hat a} \} = \{dx^{\alpha}, d\hat y^{\hat \alpha}\}$ be the coordinate basis (in special coordinate system constructed in previous subsection) for the tangent and cotangent spaces $T_{\hat z}(T \cM)$ and $T^*_{\hat z}(T \cM)$ respectively at $\hat z$ in $U_{T \cM}$. A rank $(m, n)$ tensor is given by,
\bea
\hat t &=& \hat t^{\hat a_1 \cdots \hat a_m}{}_{\hat b_1 \cdots \hat b_n} (x, \hat y) \hat e_{\hat a_1} \cdots \hat e_{\hat a_m} d\hat z^{\hat b_1} \cdots d\hat z^{\hat b_n}~.
\label{tensorTM}
\eea
The tubular expansion of the components take the following form,
\bea
\hat t^{\hat a_1 \cdots \hat a_m}{}_{\hat b_1 \cdots \hat b_n} (x, \hat y) &=& \sum_{p\geq 0} \bar{\hat t}_p{}^{\,\hat a_1 \cdots \hat a_m}{}_{\hat b_1 \cdots \hat b_n \hat \dlt_1 \cdots \hat \dlt_p} (x) \hat y^{\hat \dlt_1} \cdots \hat y^{\hat \dlt_p}~,
\eea
where $\bar{\hat t}_p{}^{\,\hat a_1 \cdots \hat a_m}{}_{\hat b_1 \cdots \hat b_n \hat \dlt_1 \cdots \hat \dlt_p}(x) $ are expressed in terms of geometric quantities of $\cM$ evaluated at $x$. The coordinate of a point in $U_{\cLM}$ is given by eq.(\ref{FNC-LM}). The coordinate basis for the tangent and cotangent spaces $T_z(\cLM)$ and $T^*_z(\cLM)$
are given by $\{e_a \} = \{e_{\alpha} = {\del \over \del x^{\alpha}}, e_A \}$ and $\{dz^a \} = \{dx^{\alpha}, dy^A \}$ respectively, where,
\bea
e_A = \oint e^{i \ha \s} {\dlt \over \dlt \hat Y^{\hat \alpha}(\s)} ~, \quad dy^A = \oint e^{-i \ha \s} \dlt \hat Y^{\hat \alpha}(\s)~,
\eea
$\dlt \hat Y^{\hat \alpha}(\sigma)$ being a functional differential. The tensor corresponding to that in (\ref{tensorTM}) is given by\footnote{The definition (\ref{tensor-LM}, \ref{tensor-comp-LM}) is equivalent to the following alternative expression
\bea
t &=& \oint t^{\hat a_1 \cdots \hat a_m}{}_{\hat b_1 \cdots \hat b_n} (x, Y(\s)) {\dlt \over \dlt \hat z^{\hat a_1}(\s)} 
\cdots 
{\dlt \over \dlt \hat z^{\hat a_m}(\s)} d\hat z^{\hat b_1}(\s) \cdots d\hat z^{\hat b_n}(\s)~,
\eea
where $\hat z^{\hat a}(\s) = (x^{\alpha}, \hat Y^{\hat \alpha}(\s))$. },
\bea
t &=& t^{a_1 \cdots a_m}{}_{b_1 \cdots b_n} (x, y) e_{a_1} \cdots e_{a_m} dz^{b_1} \cdots dz^{b_n}~,
\label{tensor-LM}
\eea
where
\bea
t^{a_1 \cdots a_m}{}_{b_1 \cdots b_n} (x, y) &=& \oint \hat t^{\hat a_1 \cdots \hat a_m}{}_{\hat b_1 \cdots \hat b_n} (x, \hat Y(\s) ) e^{-i(\ha_1 +\cdots + \ha_m)\s + i(\hb_1 + \cdots + \hb_n)\s}~.
\label{tensor-comp-LM}
\eea
Similar expression holds for the tubular expansion,
\bea
t^{a_1 \cdots a_m}{}_{b_1 \cdots b_n} (x, y) &=& \sum_{p\geq 0} \bar t_p{}^{a_1 \cdots a_m}{}_{b_1 \cdots b_n D_1 \cdots D_p} (x) y^{D_1} \cdots y^{D_p}~,
\eea
where,
\bea
\bar t_p{}^{a_1 \cdots a_m}{}_{b_1 \cdots b_n D_1 \cdots D_p} (x) &=& \oint \bar{\hat t}_p{}^{\,\hat a_1 \cdots \hat a_m}{}_{\hat b_1 \cdots \hat b_n \hat \dlt_1 \cdots \hat \dlt_p}(x)  e^{-i(\ha_1 + \cdots + \ha_m)\s + i(\hb_1+\cdots +\hb_n + \hd_1 + \cdots + \hd_p)\s}~, \cr && \cr
&=& \bar{\hat t}_p{}^{\,\hat a_1 \cdots \hat a_m}{}_{\hat b_1 \cdots \hat b_n \hat \dlt_1 \cdots \hat \dlt_p}(x) ~\dlt_{-\ha_1 - \cdots - \ha_m, + \hb_1+\cdots +\hb_n + \hd_1 + \cdots + \hd_p, 0}~.
\eea

Notice that the notations adopted in this subsection for the FNC and tensors of $\cLM$ are same as that for the FNC and tensors of $L$ considered in \S \ref{s:FNC}. This makes the expressions for the tubular expansion of various tensors of $\cLM$ look exactly the same as the corresponding general expressions in the finite dimensional case with the only additional input that the expansion coefficients are related to certain geometric data of $\cM$. Such relations are inherited from their counterparts in $T \cM$ through the above framework which relates tensors of $\cLM$ to those of $T \cM$. In particular, one finds that the expansions of the loop-space-metric-components are given by the same equations as in (\ref{g-expansion}) with the following identifications,
\bea
\eta_{A B} &=& \eta_{(\hat \alpha \hat \beta)} \dlt_{\ha + \hb, 0}~, \cr
\bar \omega_{\alpha \beta D} &=& 0 ~, \quad (\Rightarrow \bar s_{\alpha \beta D} = 0)~, \cr
\bar \omega_{\alpha B D} &=& E_{(\hat \beta) \hat \gamma}(x) \del_{\alpha}E_{(\hat \delta)}{}^{\hat \gamma}(x) 
\dlt_{\hb + \hd, 0}~, \cr
\bar r_{\alpha D E \beta} &=& 2q \check R_{\alpha (\hat \delta \hat \eta) \beta}(x) \dlt_{\hd + \hbox{e}, 0}~,\cr
\bar r_{\alpha B D E} &=& 0~, \cr
\bar r_{A B D E} &=& \check R_{(\hat \alpha \hat \beta \hat \dlt \hat \eta)}(x) \dlt_{\ha + \hb + \hd + \he, 0}~.
\label{tub-coeff-LM}
\eea
In particular, the above shows that $\cM \hookrightarrow \cLM$ is a totally geodesic submanifold. See also comments below eq.(\ref{submanifold-eqn}). With this we end our discussion of the explicit construction of the tubular neighborhood of $\cM \hookrightarrow \cLM$ and the relevant FNC. 

\section{Analogy with finite dimensional model}
\label{s:analogy}

In \S \ref{s:finite-dim} we discussed a finite-dimensional analogue of LSQM. The primary goal of this analysis was to work out the relevant details (that one would eventually like to understand for LSQM) in a finite-dimensional set up which is free of divergences. Here we will spell out how precisely to interpret the analysis of \S \ref{s:finite-dim} in the context of string theory. Our final goal will be to understand the features of the expected tachyon effective equation the analogue of which is described by eqs.(\ref{H-Heff}, \ref{Heff-comp}, \ref{tachyonEOM}).

\begin{itemize}
\item
The ambient space $L$ of the toy model is the configuration space and therefore is considered to be of Euclidean signature in \S \ref{s:finite-dim}. $\cM$, on the other hand, is the extended configuration space (which includes time) of NLSM. In the context of LSQM the analysis of \S \ref{s:finite-dim} should be viewed as a worldline type theory. This has the following consequences.
\begin{itemize}
\item
The theory is supplemented with the standard ghost sector of bosonic string theory. When $\cM$ is taken to be pseudo-Riemannain, the potential $V$ (to be discussed further below) of the model will be maximized, instead of being minimized, on the submanifold along the time-like directions. This gives rise to the standard problem of negative norm states which is cured by the presence of ghost sector. With this understanding we will simply ignore this problem now onwards and assume $\cM$ to have Euclidean signature. 

\item
Hamiltonian is a constraint. The effective form of this constraint on the submanifold obtained by integrating out the transverse (internal) degrees of freedom is supposed to give the linearized equation of motion for the string field components on $\cM$. This explains why eq.(\ref{tachyonEOM}) has been interpreted to be analogue of the linearized tachyon effective equation.
\end{itemize}

\item
In the context of finite dimension in \S \ref{s:FNC} we followed certain notation and convention for coordinate indices, FNC, tensors and their tubular expansion (see first few paragraphs of \S \ref{s:FNC}). We followed the same rules in the context of $\cLM$ in \S \ref{sss:looping}. The prescription for translating any finite dimensional expression involving tubular expansion is simply to interpret the transverse indices (i.e. capital Latin indices) according to rules of loop space as described below eq.(\ref{GC-LM}) and evaluate the barred quantities involved in terms of the intrinsic data of $\cM$ following eqs.(\ref{tub-coeff-LM}). 

\item
We now explain the potential of the toy model. Equation (\ref{vanishing-omega}) is simply the second equation in (\ref{tub-coeff-LM}). This implies that the submanifold is totally geodesic\footnote{In the context of $\cLM$, as pointed out below eq.(\ref{submanifold-eqn}), this is also related to the fact that the submanifold is the fixed-point set of the reparametrization Killing vector $v(z)$ in (\ref{v-def}). Such a feature, however, is not shared by the toy model.}.

The relevance of eq.(\ref{pot}) may be understood as follows. The potential $V^{\cLM}$ of LSQM, given in (\ref{LSCM}), can be written in terms of FNC as follows,
\bea
V^{\cLM} &=& {1\over 2} \sum_{\ha, \hb} (-) \ha \hb g_{A B} y^A y^B~, \cr
&=& {1\over 2} \sum_{\ha, \hb} ( -\ha \hb \eta_{AB} y^A y^B - {1\over 3} \ha \hb \bar r_{A C D B} y^A y^B y^C y^D + \cdots)~,
\label{V-LM}
\eea
where in the second line we have used the metric-expansion up to quadratic order from eqs.(\ref{g-expansion}). We perform the similar expansion in eq.(\ref{pot}) and then compare with the loop space potential in (\ref{V-LM}) up to quartic order. At the quadratic order one finds, using the first equation in (\ref{tub-coeff-LM}),
\bea
\eps_A = |\ha |~.
\label{epsA}
\eea 
With this identification, however, the terms at the quartic order fail to be equal as the relevant term in $V^{\cLM}$ is sensitive to the sign of the integer-pre-factor $\ha \hb$.  This changes the coefficient of the last term of $V^{\bar r}_{eff}(x)$ in (\ref{Heff-comp}) \footnote{Note that rescaling of LSQM is defined by the same eqs.(\ref{H-def}, \ref{y-rescaling}) with the identification (\ref{epsA}). }. We will see that this does not affect our final conclusion about the tachyon effective equation.

Finally, unlike the toy model, the loop space potential $V^{\cLM}$ is invariant under the full GCT (see appendix \ref{a:note}) of $\cLM$. This happens because of the special property of loop space that it admits a vector field that is linear in coordinate. Therefore, the general covariance of LSQM is broken down to that of the submanifold only in the semi-classical vacuum, in particular, because of the rescaling (\ref{y-rescaling}). 

\item
The leading order rescaled Hamiltonian given by the first equation in (\ref{epsH012}) is analogous to the non-zero mode contribution to the Hamiltonian in flat space. The resemblance can be made more explicit through the following redefinition:
\bea
\lt. \begin{array}{ll}
a^A \to \alpha^A ~, &  a^{\dagger}{}^{\bar A} \to \alpha^{\bar A} ~, \cr
a^{\bar A} \to \tilde \alpha^A ~, & a^{\dagger}{}^A \to \tilde \alpha^{\bar A} ~, 
\end{array} \rt\} \quad \ha >0~,
\label{alpha-oscillators}
\eea
where $\alpha$ and $\tilde \alpha$ are the usual flat-space-oscillators \cite{polchinski}. The index $\bar A \to (\alpha, -\ha)$ (see discussion below eq.(\ref{GC-LM})) corresponds to a negative mode while $A$ to a positive mode. The leading order Hamiltonian takes the following familiar form in this new notation,
\bea
{}^{\eps}\cH^{(0)} &=&\sum_{\ha, \hb >0} \sqrt{\ha \hb} ~\eta_{\bar A B} (\alpha^{\bar A} \alpha^B + \tilde \alpha^{\bar A} \tilde \alpha^B) + \sum_{\ha>0} \ha ~,
\eea
The last term is the zero-point energy which is divergent. Noticing that at the leading order the transverse dynamics exactly matches with that of the non-zero modes in flat space, this term can be treated in the usual manner. The point to be emphasized here is that in flat space such a term (after collecting the ghost contribution) finally gets related to the tachyon mass.  
The same is true here as we see from the first equation in (\ref{Heff-comp}). Notice also that this mass has the right scaling with respect to $\hbar=\alpha'$. In fact, demanding that the leading order transverse Hamiltonian in (\ref{epsH012}) be precisely same as the non-zero modes contribution to the Hamiltonian in flat space fixes the rescaling of the model as described by eqs.(\ref{H-def}, \ref{y-rescaling}). Such a condition is required to get the right flat space limit (where the tubular expansion becomes trivial) of our analysis.

\item
Notice that while reinterpreting the oscillators of the toy model in terms of string modes through eqs.(\ref{alpha-oscillators}), the creation and annihilation operators do not mix up. This implies that the results for the vacuum expectation values in eqs.(\ref{VEVs}) are also valid in the context of LSQM. This enables us to view eqs.(\ref{H-Heff}, \ref{Heff-comp}, \ref{tachyonEOM}) as describing the linearized tachyon effective equation at leading order in LSQM. One expects that such an equation should be given by,
\bea
(-{1\over 2} \nabla^2 + {m^2\over 2}) T(x) = 0~,
\label{tachyonEOM2}
\eea
up to leading order equation of motion (i.e Ricci flatness) for the background,
\bea
R_{\alpha \beta}(x) = 0~.
\label{Ricci-flatness}
\eea
This equation has not been derived in this work as the toy model does not have an analogue of conformal invariance. However, one can imagine deriving this condition at the leading order of tubular expansion of the DeWitt-Virasoro algebra computed in \cite{dwv}. We postpone this analysis for a future work and assume this is true for the time being.

We will now argue that,
\bea 
V^{\bar r}_{eff}(x) &\propto & R(x)~, \cr
V^{\bar \omega}_{eff}(x) &=& 0~,
\eea
where the proportionality constant in the first equation is divergent and this is the only divergence in the effective equation of motion at leading order. This is established simply by using the relevant equations in (\ref{tub-coeff-LM}). Notice that the undetermined factor $q$ does not influence our final conclusion which is insensitive to the proportionality constant (it is divergent anyway). To establish the second equation one notices from the third equation in (\ref{tub-coeff-LM}) that $\bar \omega_{\alpha B D}$ is non-zero only when $\hb + \hd =0$, in which case $\eps_B=\eps_D$ (according to the identification in (\ref{epsA})), and therefore $V^{\bar \omega}_{eff}(x)$ in eqs.(\ref{Heff-comp}) vanishes.

\end{itemize}

With this we end our discussion of how the computations in the finite-dimensional model discussed in \S \ref{s:finite-dim} should be interpreted in the context of LSQM.

\section{Conclusion}
\label{s:conclusion}

This work investigates how to make sense of a semi-classical limit of LSQM as discussed in \cite{dwv}. In this limit the wavefunction gets localized on the submanifold $\cM$ of vanishing loops in $\cLM$ where $\cM$ is the target space of the corresponding NLSM. The study involves first defining the procedure in a finite dimensional toy model (content of \S \ref{s:FNC} and \S \ref{s:finite-dim}) and then figuring out how the actual loop space model can be understood through an analogy with the toy model (content of \S \ref{s:tubular} and \S \ref{s:analogy}). The study shows that the linearized effective equation for the tachyon fluctuation at leading order in $\alpha'$-expansion is reproduced correctly with all the divergent terms being proportional to the Ricci scalar of $\cM$. 

The present approach makes the usual picture of particle quantum mechanics quite explicit and therefore it is conceptually appealing. Given this, it is perhaps a good idea to work out the details of how the standard questions, such as Ricci-flatness as leading order condition for conformal invariance, low-energy effective equations of motion and, most importantly, higher order $\alpha'$ corrections, should be understood in the current approach. We hope that the analysis of the present work will be helpful for further study along this direction and its supersymmetrization.

We will conclude with a few remarks regarding the mathematical framework of \S \ref{ss:construct-tubular} and \S \ref{ss:FNCinLM} where a certain Riemannian structure on $T\cM$ was discussed. A speciality of this Riemannian structure is that it views $T\cM_0 \hookrightarrow T \cM$ as a submanifold admitting a tubular neighborhood. Recall that an all-order understanding of tubular expansion of vielbein in a generic case is available through \cite{tubular} (reviewed in appendix \ref{a:vielbein}). This implies that finding the desired Riemannian structure on $T\cM$ is equivalent to finding all the tubular expansion coefficients in terms of intrinsic geometric data of $\cM$. In this work this has been done in a limited sense which proved sufficient for the present level of analysis of  LSQM. It is possible that a more complete understanding of this question will be required for computing $\alpha'$ corrections. We hope to come back to these questions in future.

Finally, we note that the mathematical framework discussed in \S \ref{ss:construct-tubular} should also be relevant for a multi-particle classical system in curved space. The center of mass (CM) of the system belongs to the first copy of $\cM$ in $\cM \times \cM$ while the positions relative to CM lie in the second copy. After performing the bundle map in (\ref{bundle-map}) the CM resides in $T \cM_0$ while the relative coordinates all lie on the same fibre whose base point is identified with the position of CM. In the limit when the potential is strong enough to hold all particles in the form of a rigid body, the dynamics confines on the submanifold $T\cM_0 \hookrightarrow T \cM$. When we are away from this limit, the internal fluctuations are described in the tubular neighborhood of this embedding.

\begin{center}
{\bf Acknowledgement}
\end{center}

I am thankful to A. P. Balachandran, Partha Sarathi Chakraborty, Sumit R. Das, Ghanashyam Date, Rukmini Dey, Debashis Ghoshal, Rajesh Gopakumar, Dileep P. Jatkar, Koushik Ray, B. Sathiapalan, Ashoke Sen, Alfred D. Shapere and Andrew Stacey for useful discussion, questions and comments. I thank Harish-Chandra Research Institute (Allahabad) for hospitality and the organizers of  the meetings ``Quantum Theory and Symmetries 7'' (Prague, August 2011) and ``Indian National Strings Meeting 2011'' (New Delhi, December 2011) where preliminary versions of this work were presented. Study of \S \ref{s:finite-dim} and related matters have been facilitated by the use of the mathematical program Cadabra \cite{cadabra}. I thank Kasper Peeters for helpful communication.

\appendix

\section{All order tubular expansion of vielbein}
\label{a:vielbein}

In order to derive closed form expressions for the tubular expansion coefficients for vielbein, one starts with the following set of conditions which are equivalent to (\ref{coord}),
\bea
e^{(a)}{}_B y^B = \bar e^{(a)}{}_B y^B~, \quad y^A \omega_A{}^{(b)}{}_{(c)} = 0~,
\eea
where $e^{(a)}{}_b$ is the vielbein of $L$. Then by making use of the Cartan's structure equations (as was done in \cite{muller}) one derives the following second order differential equation \cite{tubular},
\bea
\hbd (\hbd + \mathop{\eps}^b) e^{(a)}{}_b(x,y) = \rho^{(a)}{}_{(c)}(x,y;y) e^{(c)}{}_b(x,y)~,
\label{diff-viel}
\eea
where for any function $f(x,y)$ we have defined,
\bea
\hbd f(x,y) = y^A \del_A f(x,y)~.
\label{db}
\eea
Furthermore,
\bea
\rho^{(a)}{}_{(b)} (x,y; \tilde y) \equiv r^{(a)}{}_{CD (b)}(x,y) \tilde y^C \tilde y^D~,
\eea
and 
\bea
\mathop{\eps}^b = \lt\{\begin{array}{rl}
1 & \quad \hbox{when } b=B ~, \cr
-1 & \quad \hbox{otherwise}~.
\end{array} \rt.
\label{eps}
\eea

The solution to eq.(\ref{diff-viel}) can be given in the form of the following tubular expansion \cite{tubular},
\bea
e^{(a)}{}_B &=& \sum_{n=0}^{\infty} \sum_{s_1, \cdots , s_n = 0}^{\infty} \cF^{(n)}_{\perp}(s_1,s_2, \cdots ,s_n) 
\lt[(y.\cD)^{s_1} \rho(x,0;y) \cdots (y.\cD)^{s_n}
\rho(x,0;y) \rt]^{(a)}{}_{(b)} e_0^{(b)}{}_B ~, \cr
e^{(a)}{}_{\beta} &=& \sum_{n=0}^{\infty} \sum_{s_1, \cdots , s_n = 0}^{\infty} \cF^{(n)}_{\parallel}(s_1,s_2, \cdots
,s_n) \lt[(y.\cD)^{s_1} \rho(x,0;y) \cdots (y.\cD)^{s_n}
\rho(x,0;y) \rt]^{(a)}{}_{(b)} e_0^{(b)}{}_{\beta}~, \cr  &&
\label{viel-exp}
\eea
where,
\bea
\cF_{\perp}^{(n)}(s_1,s_2, \cdots ,s_n) &=& {
C^{s_1+s_2+\cdots +s_n+2n-1}_{s_1} C^{s_2+s_3+\cdots
  +s_n+2n-3}_{s_2} \cdots C^{s_n+1}_{s_n} \over (s_1+ s_2+\cdots +s_n+ 2n+1)!}~, \cr
\cF_{\parallel}^{(n)}(s_1,s_2, \cdots ,s_n) &=& {
C^{s_1+ s_2+ \cdots +s_n+2n-2}_{s_1} C^{s_2+ s_3+ \cdots
  +s_n+2n-4}_{s_2} \cdots 1 \over (s_1+ s_2 +\cdots
  +s_n+ 2n)!}~.
\eea
$C^n_r$ are the binomial coefficients and,
\bea
[(y. \cD)^s \rho(x, 0;y)]^{(a)}{}_{(b)} &=& \cD_{A_1} \cdots \cD_{A_s} r^{(a)}{}_{C D(b)} (x, 0) y^{A_1}\cdots y^{A_s}  y^C y^D~,
\eea
where $\cD_a$ is the covariant derivative in $L$ with respect to the metric $g_{a b}$. Notice that all such derivatives are evaluated at the submanifold. Finally\footnote{Recall, according to our rule for notation, $\bar \omega_{\alpha}{}^{(b)}{}_C = \omega_{\alpha}{}^{(b)}{}_C (x, y=0)$.}, 
\bea
e_0^{(a)}{}_B &=& \dlt^{(a)}{}_B~, \cr
e_0^{(a)}{}_{\beta} &=& \lt\{ \begin{array}{ll}
E^{(\alpha)}{}_{\beta} + \bar \omega_{\beta}{}^{(\alpha)}{}_C y^C  ~, & \hbox{ for } a=\alpha ~, \\ & \\
\bar \omega_{\beta}{}^{(A)}{}_C y^C ~, & \hbox{ otherwise }~,
\end{array} \rt.
\label{e_0}
\eea
$E^{(\alpha)}{}_{\beta}(x) $ being the vielbein of the induced metric on $M$,
\bea
\bar g_{\alpha \beta} = G_{\alpha \beta}(x) = E^{(\gamma)}{}_{\alpha}(x) E_{(\gamma) \beta}(x) ~.
\eea
Using the results in (\ref{viel-exp}) we find the metric expansion given in (\ref{g-expansion}).

\section{Tubular expansion of Hamiltonian}
\label{a:tubeH}

Here we will present the detailed computations required to carry out various steps of performing semi-classical expansion as defined in \S \ref{ss:def}. The rescaling of wavefunction that takes us to the submanifold based description is given by,
\bea
\psi(z) = \lt({g\over G}\rt)^{1/4} \psi'(z)~.
\eea  
This leads to the following expression for $\cH^{sub}$ as defined in eq.(\ref{cHsub}),
\bea
\cH^{sub} &=& - {\hbar^2 \over 2} (\cD^2 + \cT + \cU) + V~,
\label{cHsub-result}
\eea
where,
\bea
\cT &=& -{1\over 2} g^{a b} l_a \del_b ~, \quad
\cU = {1\over 16} g^{a b} l_a l_b + {1\over 4} (- g^{a b} q_{a b} + \gamma^c l_c)~, \cr
l_a &=& \del_a \ln\lt({g \over G}\rt)~,  \quad 
q_{a b} = \del_a\del_b \ln\lt({g\over G}\rt)~, 
\eea
The contribution at $O(y^n)$ in the tubular expansion of $\cH^{sub}$ is given by,
\bea
\cH^{sub}_n &=& -\hbar^2 (K^{\parallel}_n + K_n + K^{\perp}_n) + V_n ~,
\label{cHsubn-result}
\eea
where,
\bea
K^{\parallel}_n &=& {1\over 2}(\mD^{\parallel}_n - d^{\parallel}_n+ t^{\parallel}_n + \cU_n) ~, \cr
K_n &=& {1\over 2} (2 \mD_n -d_n + t_n) ~, \cr
K^{\perp}_n &=& {1\over 2} \mD^{\perp}_n ~,
\label{Kn-result}
\eea
and\footnote{Given a geometric quantity $X$, the notation $X_n$ has been explained below eq.(\ref{cHsub-n}).},
\bea
\mD^{\parallel}_n &=& g_n^{\alpha \beta} \del_{\alpha} \del_{\beta} ~, \quad 
\mD_n = g_n^{\alpha B}\del_{\alpha} \del_B ~, \quad \mD^{\perp}_n = g_n^{A B}\del_A \del_B \cr
d^{\parallel}_n &=& \gamma_n^{\alpha} \del_{\alpha} ~, \quad d_n = \gamma_n^A \del_A ~, \cr
t^{\parallel}_n &=& -{1\over 2} \sum_{m=0}^{n}(g_{n-m}^{\alpha \beta} l_{m \alpha}
+ g_{n-m}^{A \beta} l_{m A}) \del_{\beta} ~, \cr
t_n &=& -{1\over 2} \sum_{m=0}^{n} (g_{n-m}^{\alpha B} l_{m \alpha}   
+ g_{n-m}^{A B} l_{m A}) \del_B ~.
\label{Dn-dn-tn-result}
\eea
Finally, the contribution at $O(\hbar^{n\over 2})$ in the semi-classical expansion of $\cH$ works out to be\footnote{Given $X$, the notation ${}^{\eps}X$ has been defined in eq.(\ref{eps-notation}).},
\bea
{}^{\eps}\cH^{(n)} &=& -{}^{\eps}K^{\parallel}_{n-2} - {}^{\eps}K_{n-1} - {}^{\eps}K^{\perp}_{n} + {}^{\eps}V_{n+2}~. 
\label{calH-n2}
\eea
This shows how the contribution at a given order in $\hbar$ is related to terms in tubular expansion of various geometric quantities at different orders.

\section{A note on loop space and LSQM}
\label{a:note}

The loop space $\cLM$ associated to a Riemannian manifold $\cM$ is the space of all smooth maps from a parametrized loop to $\cM$.
\bea
\cLM &=& C^{\infty}(S^1, \cM)~.
\label{LM-def}
\eea
Here we will briefly note down some general features of $\cLM$ and LSQM that are relevant for our discussion in this article. The above definition implies that given any element $l \in \cLM$, there exists a smooth function $p^{(l)}:S^1 \to \cM$. We wish to define a general coodinate system in $\cL \cM$ in an open neighborhood of small loops as defined in \S \ref{ss:construct-tubular}. To this end we recall the definition of all the open sets as given below eq.(\ref{loop-ave-TxM}). Therefore, given any $l\in C(U_{\cM})$, the entire image $p^{(l)}$ resides inside $\tilde U_{\cM}$. Let $x: \tilde U_{\cM} \to \mR^D$ be a local coordinate system in $\tilde U_{\cM} \subset \cM$. Then $Z^{(l)}= x \circ {p^{(l)}}: S^1 \to \mR^D$, and therefore is an element of  ${\cal L}\mR^D$ which is the model space of $\cLM$. We consider $Z: C(U_{\cM}) \to {\cal L}\mR^D$, such that $Z\circ l = Z^{(l)} \in {\cal L}\mR^D$, to be the coordinate functions in the relevant neighborhood in $\cLM$. 

Following \cite{dwv}, we will work with a Fourier space representation of these coordinate functions. In this representation the general coordinates of a point $l$ as considered above in $\cLM$ are given by,
\bea
z^a &=& \oint Z^{\alpha}(\s) e^{-i \ha \s} ~, \quad \ha \in \mZ~.
\label{GC-LM}
\eea
where $\s$ parametrizes the loop, $\oint \equiv \int {d\s \over 2\pi}$ and the loop embedding $Z^{\alpha}(\s)$ ($\alpha$ being a target space index) is obtained by following the above definition. We adopt the following convention for an infnite-dimensional coordinate index. It is given by a lower case Latin alphabet, which in turn is associated to a pair containing the corresponding Greek alphabet (i.e. a target space index) and an integer, denoted by the same small Latin alphabet in text format. For example, $a \to (\alpha, \ha)$, $b \to (\beta, \hb)$. We will also adopt a similar association between such a pair and the corresponding upper case Latin alphabet when the integer is non-zero, i.e. $A \to (\alpha, \ha)$, $B \to (\beta, \hb)$ etc. only when $\ha, \hb \neq 0$. We use this type of notation in all our discussion involving an explicit coordinate system in $\cL \cM$.

Reparametrization of the loop corresponds to an isometry of the loop space which exists irrespective of the isometries of $\cM$. The generator of this isometry is given by,
\bea
v^a(z) &=& \oint \del_{\sigma} Z^{\alpha}(\sigma) e^{-i \ha \s}~.
\label{v-def}
\eea
which satisfies the Killing vector equation in $\cLM$ \cite{schreiber}.
\bea
&& \cD_a v_b + \cD_b v_a = 0~, 
\label{v-isometry}
\eea
where $\cD_a$ is the covariant derivative on $\cLM$. The metric and affine connection on $\cLM$ are given by,
\bea
g_{a b}(z) &=& \oint G_{\alpha \beta}(Z(\s)) e^{i(\ha + \hb)\s} ~,  \cr
\gamma^a_{b d}(z) &=& \oint \Gamma^{\alpha}_{\beta \delta}(Z(\sigma)) e^{i(-\ha + \hb + \hd) \sigma}~,
\eea
respectively, where $G_{\alpha \beta}$ and $\Gamma^{\alpha}_{\beta \delta}$ are the metric and affine connection on $\cM$ respectively. Notice from eqs.(\ref{v-def}) that the submanifold of vanishing loops, whcih is given by,
\bea
z^A = \oint Z^{\alpha}(\sigma) e^{-i\ha \s} = 0~, \quad \forall \ha \neq 0~,
\label{submanifold-eqn}
\eea
is where the Killing vector field vanishes. This situation is similar to the consideration of Kobayashi's theorem in \cite{kobayashi} (in finite dimensions), which claims that the space of fixed points of an isometry is a totally geodesic submanifold of even co-dimension. We will see in \S \ref{s:tubular} that the submanifold of interest is indeed totally geodesic. Although, this has infinite number of transverse directions, from the discussion of the infinite dimensional coordinate index done below eq.(\ref{GC-LM}), it is clear that for every transverse index $A \to (\alpha, \ha)$, there is a pair $\bar A \to (\alpha, -\ha)$.  

We now briefly recall the structure of LSQM following \cite{dwv}. The classical NLSM Lagrangian on a flat worldsheet takes the following form in terms of the general coordinates in $\cLM$,
\bea
L &=& K - V^{\cLM}~, \cr
K &=& {1\over 2} g_{a b}(z) \dot z^a \dot z^b ~, \cr
V^{\cLM} &=& {1\over 2} g_{a b}(z) v^a(z) v^b(z)~,
\label{LSCM}
\eea
where a dot indicates derivative with respect to the worldsheet time. Notice that the potential is proportional to the norm-square of the Killing vector field discussed above. LSQM \cite{dwv} is a formal $\hbar$-deformation of this classical system obtained by following DeWitt's argument in \cite{dewitt}. Therefore, it has the same mathematical structure as that of the toy model discussed in \S \ref{s:finite-dim} with the configuration space replaced by the infinite dimensional loop space\footnote{Though a crucial difference is that LSQM should be viewed as a worldline-type description of a relativistic system. See \S \ref{s:analogy} for more details on this.}. In particular, the matrix element of the Hamiltonian between two scalar states is given by the same equation as in (\ref{Hpre}), with various quantities now interpreted in the context of $\cLM$ instead of $L$.  For example, $\cD^2$ denotes the Laplacian in $\cLM$.

\section{Existence of $(x', y')$-system}
\label{a:existence}

In the discussion of \S \ref{sss:TM} we assumed that starting from the direct product coordinate system $\bar z=(x_1, x_2)$ (see eqs.(\ref{z-bar}, \ref{ebar-comp})) on $\cM \times \cM$ one can arrive at another, namely $z'=(x', y')$, such that the transformed vielbein components are given, up to a constant conformal transformation, by eqs.(\ref{arbit-prime}) up to quadratic order in $y'$. 
Here we will explicitly construct $z'$ in a region whose overlap with the diagonal submanifold is sufficiently small.

We begin by discussing geodesics of $\cM \times \cM$. These are direct product of geodesics in the two copies of $\cM$. The ones which pass through $\bar z_0 = (x_0, x_0) \in \Dlt_{U_{\cM \times \cM}} (=  U_{\cM \times \cM} \cap \Dlt)$ can be labelled by a unit vector $({1\over \sqrt{2}}\eta, {1\over \sqrt{2}} \zeta) \in T_{x_0}(\cM) \times T_{x_0}(\cM)$, where $\eta^{\alpha_1}$ ($\zeta^{\alpha_2}$) is the unit tangent to the geodesic in the first copy (second copy) at $x_1=x_0$ ($x_2=x_0$). The vectors $({1\over \sqrt{2}}\eta, {1\over \sqrt{2}} \eta)$ and $({1\over \sqrt{2}}\eta, -{1\over \sqrt{2}}\eta)$ are parallel and transverse to $\Dlt_{U_{\cM \times \cM}}$ at $(x_0, x_0)$ respectively. A geodesic whose unit tangent vector at $(x_0, x_0)$ is of the form $({1\over \sqrt{2}}\eta, {1\over \sqrt{2}} \eta)$ remains on the diagonal submanifold for ever. This implies that $\Delta$ is a totally geodesic submanifold of $\cM \times \cM$. We would like to construct $z'$ such that $y'$ is a geodesic coordinate along the transverse direction given by a unit vector of the form $({1\over \sqrt{2}}\eta, -{1\over \sqrt{2}}\eta)$.

We first consider the following coordinate transformation,
\bea
\bar z^a \to \tilde z^a = (y_1^{\alpha_1}, y_2^{\alpha_2})~,
\eea
such that,
\bea
x_1^{\alpha_1} = \exp_{x_0}^{\alpha_1}(y_1)~, \quad x_2^{\alpha_2} = \exp_{x_0}^{\alpha_2}(y_2)~,
\eea
where the exponential map $\exp_{x_0}: T_{x_0}\cM \to \cM$ is given by,
\bea
\exp^{\alpha}_{x_0}(\xi) &=& x_0^{\alpha} + \xi^{\alpha} - {1\over 2} \Gamma^{\alpha}_{\beta_1 \beta_2}(x_0) \xi^{\beta_1} \xi^{\beta_2} + \cdots ~,
\label{exp-map}
\eea
$\Gamma^{\alpha}_{\beta_1 \beta_2}$ being the Christoffel symbols of $\cM$. We readily recognize that $y_1$ and $y_2$ are individually $\hbox{RNC}_{x_0}$ in the two copies of $\cM$ and the system $\tilde z$ is an $\hbox{RNC}_{(x_0, x_0)}$ in $\cM \times \cM$. The transformed vielbein components are expanded up to quadratic order as \cite{muller},
\bea
\tilde e^{[\alpha_1]}{}_{\beta_1} (\tilde z) &=& E^{(\alpha_1)}{}_{\beta_1}(x_0) + {1\over 6} \check R^{(\alpha_1)}{}_{\gamma_1 \delta_1 \beta_1}(x_0) y_1^{\gamma_1} y_1^{\delta_1}~, \cr
\tilde e^{[\alpha_1]}{}_{\beta_2} (\tilde z) &=& 0 ~, \cr
\tilde e^{[\alpha_2]}{}_{\beta_1} (\tilde z) &=& 0 ~, \cr
\tilde e^{[\alpha_2]}{}_{\beta_2} (\tilde z) &=& E^{(\alpha_2)}{}_{\beta_2}(x_0) + {1\over 6} \check R^{(\alpha_2)}{}_{\gamma_2 \delta_2 \beta_2}(x_0) y_2^{\gamma_2} y_2^{\delta_2}~.
\eea

Next we define the orthonormal frames which are parallel and transverse to the diagonal by giving the following rotation in the local Lorentz frame,
\bea
\tilde e^{[\alpha]}{}_b = {1\over \sqrt{2}} (\tilde e^{[\alpha_1=\alpha]}{}_b + \tilde e^{[\alpha_2=\alpha]}{}_b)~,\cr
\tilde e^{[\hat \alpha]}{}_b = {1\over \sqrt{2}} (\tilde e^{[\alpha_1= \hat \alpha]}{}_b - \tilde e^{[\alpha_2=\hat \alpha]}{}_b)~,
\label{local-rotation}
\eea
where $\alpha, ~\hat \alpha = 1, \cdots, D$. The parallel and transverse coordinates, $u^{\alpha}$ and $y''^{\hat \alpha}$ respectively, are defined by yet another coordinate transformation,
\bea
u^{\alpha} = {1\over 2} (y_1^{\alpha_1=\alpha} + y_2^{\alpha_2=\alpha})~, \quad y''^{\hat \alpha} = {1\over 2} (y_1^{\alpha_1=\hat \alpha} - y_2^{\alpha_2=\hat \alpha})~.
\eea
Therefore $u^{\alpha}$ is an $\hbox{RNC}_{(x_0, x_0)}$ of $\cM \times \cM$ which remains parallel to the submanifold 
$\Delta$.  We combine $u^{\alpha}$ with $x_0$ to construct a general coordinate $x''^{\alpha}$ on $\Delta$,
\bea
x''^{\alpha} = \exp_{x_0}^{\alpha}(u)~.
\eea
Therefore, we seem to have arrived at a coordinate system $z''^a = (x''^{\alpha}, y''^{\hat \alpha})$ where $x''^{\alpha}$ is a general coordinate system on $\Delta$ and $y''^{\hat \alpha}$ is orthogonal to it. However, it has been constructed using exponential map with a fixed base point. Therefore, it is guranteed to be the right one, i.e. the one relevant to FNC, only near the base point $(x_0, x_0)$. Now onwards we restrict to a region around this point whose overlap with $\Delta$ is sufficiently small. More precisely, we consider $u$ to be at higher order in smallness with respect to $y''$, implying that we neglect terms of order $u y''$ and $u^2$ with respect to those of order $y''^2$. With this approximation the transformed vielbein components in $z''$-system are given by,
\bea
{1\over \sqrt{2}} e''^{[\alpha]}{}_{\beta} (z'') &=& E^{(\alpha)}{}_{\beta}(x'') + {1\over 6} \check R^{(\alpha)}{}_{\hat \gamma \hat \delta \beta}(x'') y''^{\hat \gamma} y''^{\hat \delta}~, \cr
{1\over \sqrt{2}} e''^{[\alpha]}{}_{\hat \beta} (z'') &=& 0~, \cr
{1\over \sqrt{2}} e''^{[\hat \alpha]}{}_{\beta} (z'') &=& 0~, \cr
{1\over \sqrt{2}} e''^{[\hat \alpha]}{}_{\hat \beta} (z'') &=& E^{(\alpha)}{}_{\hat \beta}(x'') + {1\over 6} \check R^{(\alpha)}{}_{\hat \gamma \hat \delta \hat \beta}(x'')  y''^{\hat \gamma} y''^{\hat \delta}~.
\label{vielbein''}
\eea
The ${1\over \sqrt{2}}$ factors arise because of the standard constant rescaling of the measure when we go to a diagonal. Now onwards, we will absorb this by applying a constant conformal transformation of the metric.

There is a further coordinate transformation which keeps the form of the expansions in (\ref{vielbein''}) invariant within the same region of validity, yet making it more general. This is given by $z'' \to z' = (x'^{\alpha}, y'^{\hat \alpha})$ such that,
\bea
x''^{\alpha} = x'^{\alpha} + (q-{1\over 6}) \check R^{\alpha}{}_{\hat \gamma \hat \delta \beta}(x') y'^{\hat \gamma} y'^{\hat \delta} x'^{\beta} ~, \quad y''^{\hat \alpha} = y'^{\hat \alpha}~,
\eea
where $q$ is a real constant. The transformed vielbein components are given by eqs.(\ref{arbit-prime}).

\end{document}